\documentclass[12]{iopart}

\usepackage{amssymb}
\usepackage{graphicx}
\usepackage{bm}

\begin{document}

\title[Comparative pressure measurements on $\kappa$-(BEDT-TTF)$_{2}$Cu(SCN)$_{2}$]
{Comparative magnetotransport and T$_{\rm c}$ measurements\\
  on $\kappa$-(BEDT-TTF)$_{2}$Cu(SCN)$_{2}$ under pressure.}

\author{A-K Klehe\dag 
\footnote[3]{To whom correspondence should be addressed}, T Biggs\dag, C A Kuntscher\dag 
\footnote[4]{Permanent Address: 1. Physikalisches Institut, Universit\"{a}t 
Stuttgart, Pfaffenwaldring 57, D-70550 Stuttgart, Germany}, A M Kini\ddag\ and J A Schlueter\ddag\
}

\address{\dag\ Clarendon Laboratory, Department of Physics, Oxford University,\\
Parks Road, Oxford OX1 3PU, U.K.}

\address{\ddag\ Materials Science Division, Argonne National Laboratory, 9700 South
Cass Avenue, Argonne, Illinois 60439, U.S.
}%

\ead{a.klehe1@physics.ox.ac.uk}

\begin{abstract}

We compare magnetotransport measurements under pressure
on the organic superconductor
$\kappa$-(BEDT-TTF)$_{2}$Cu(SCN)$_{2}$ with different pressure-media and
discover that the results are pressure media dependent. This 
pressure-medium dependence is thought to originate from the difference in 
thermal contraction between the very soft and highly anisotropic sample and the
isotropically contracting, but solid pressure medium, thus resulting in 
non-hydrostatic pressure on the sample.
However, comparison of pressure measurements with different media
reveals 
a pressure-medium
independent correlation  
between the superconducting transition temperature, T$_{\rm c}$, and the size of the 
quasi 2-dimensional Fermi surface pocket and thus the quasi 2-dimensional 
carrier density in $\kappa$-(BEDT-TTF)$_{2}$Cu(SCN)$_{2}$. The
observed pressure-induced increase in the quasi 2-dimensional carrier density can be 
interpreted as a transfer of carriers from quasi 1-dimensional Fermi 
surface sections, 
reminiscent of a mechanism in cuprate superconductors, where pressure is known to transfer 
carriers from the insulating charge reservoir layers into the 
conducting cuprate sheets.     
\end{abstract}
\submitto{\JPCM}

\maketitle

\section{\label{sec:level1}Introduction}

$\kappa$-(BEDT-TTF)$_{2}$Cu(SCN)$_{2}$ is one of the best 
characterized organic superconductors \cite{yamaji,john1,john2}. The resemblance
of its pressure-temperature phase diagram \cite{lefebvre} to that of the 
carrier-density-temperature phase diagram in cuprate superconductors 
\cite{presland} has
frequently been taken as evidence for similar interaction mechanisms
governing superconductivity \cite{schmalian,mckenzie}. Magnetic ordering, 
metallic behaviour and 
superconductivity in cuprate superconductors are determined by
the carrier concentration
in the conducting cuprate layers \cite{presland}. This carrier concentration 
can be regulated
either through chemical doping \cite{presland} or pressure \cite{neumeier,klehe2}. 
In organic superconductors, however, evidence for a correlation between carrier 
density and the superconducting transition temperature, T$_{\rm c}$, has been missing.

$\kappa $-(BEDT-TTF)$_{2}$Cu(SCN)$_{2}$ is a typical example of a quasi 2-dimensional
superconductor. It  
is a strongly anisotropic material
in which conducting layers of (BEDT-TTF)$_{2}^{+1}$ in
the crystallographic \textit{bc}-plane are separated by insulating
layers of polymorphic Cu(SCN)$_{2}^{-1 }$ \cite{yamaji,john1,john2}. The
resulting strong anisotropy is reflected in all physical properties of the
material, including its electrical conductivity 
\cite{yamaji,john1,john2}, its quasi 2-dimensional bandstructure 
\cite{john1,john2},
its compressibility \cite{rahal}, its thermal contraction \cite{schultz1} 
and its uniaxial
pressure dependence of T$_{\mathrm{c}}$ \cite{mueller}.

Also the Fermi surface of $\kappa$-(BEDT-TTF)$_{2}$Cu(SCN)$_{2}$ reflects the 
structural anisotropy; it consists of a 
quasi 2-dimensional (Q2D) Fermi surface pocket and two quasi 1-dimensional (Q1D) Fermi
surface sections \cite{yamaji,john1,john2}. The Q2D Fermi surface pocket, also known as 
the $\alpha$-pocket, gives rise to Shubnikov-de Haas oscillations of $\sim$ 600 T
at ambient pressure. At high magnetic fields and low temperatures, magnetic 
breakdown between the
$\alpha$-pocket and the Q1D Fermi surface sections can give 
rise to a semiclassical orbit, the $\beta$-orbit, which has the same cross 
sectional area as the
Brillouin zone \cite{john1,john2}. Thus, knowledge of this orbital size is a measure of the 
low temperature in-plane compressibility of the material. In conjunction with
the size of the $\alpha$-orbit, it also allows one to calculate the exact shape
of the Q2D Fermi surface according to the effective dimer model 
\cite{john3}. 
Superconductivity in $\kappa $-(BEDT-TTF)$_{2}$Cu(SCN)$_{2}$ is thought
to be strongly affected by the detailed warping of the Fermi surface 
\cite{schmalian,Louati},
with the pairing of the electrons in the Q2D band being 
mediated by an exchange of spin fluctuations within the Q1D 
band \cite{Louati}. A highly critical parameter for the 
material is the ratio of the interdimer transfer integrals t$_{\rm b}$/t$_{\rm c}$ 
\cite{Louati}, which has been predicted \cite{Louati} and 
observed \cite{tim} to increase under pressure. This pressure induced increase
is currently considered to be the driving force of the different pressure-induced
phase transitions observed in $\kappa $-(BEDT-TTF)$_{2}$Cu(SCN)$_{2}$ \cite{Louati}. 
Recent studies \cite{john3,tim} have also discussed and
investigated the importance of the out-of plane direction, the
crystallographic \textit{a}$'$-direction perpendicular to the 
\textit{bc}-plane, for a more general understanding of the general physical 
properties of $\kappa $-(BEDT-TTF)$_{2}$Cu(SCN)$_{2}$. The interlayer transfer 
integral, t$_{\bot} \approx 0.04$ meV, was found to be a factor $\leq$ 10$^{3}$ smaller than
that observed for the intralayer components, t$_{\rm b}$ and t$_{\rm c}$ 
\cite{john3},
indicating that $\kappa $-(BEDT-TTF)$_{2}$Cu(SCN)$_{2}$ is predominantly 
Q2D in its electronic properties, even though there is a small 
degree of coherent interlayer transport \cite{john3}. The warping of the Fermi
surface in the interplane direction is so small \cite{john3} that the Landau
quantization
is almost entirely determined by the field perpendicular to the 2-dimensional
planes,
thus allowing for an easy correction of the measured Shubnikov-de Haas 
frequencies in a 
tilted magnetic field \cite{john1}.
Further evidence for the   
predominantly Q2D character of
$\kappa $-(BEDT-TTF)$_{2}$Cu(SCN)$_{2}$ is the absence of "neck-and-belly" 
frequencies in the de Haas-van Alphen effect \cite{sasaki} and the 
intrinsically broad width of the superconducting-to-normal 
transition in zero-field resistance measurements \cite{john-tc}.
   
Uniaxial pressure measurements \cite{campos} and calculations \cite{campos2} 
have been performed on $\kappa $-(BEDT-TTF)$_{2}$Cu(SCN)$_{2}$, indicating a 
strongly anisotropic pressure dependence of its physical properties. The 
observed increase of the Q2D Fermi surface area under 
pressure \cite{tim,caulfield1,caulfield2} 
as well as under uniaxial stress in the out-of-plane direction \cite{campos}, 
is thought to originate
from a pressure-induced intradimer sliding motion \cite{campos2}.
 
The value of T$_{\rm c}$ exhibits
an inverse isotope effect upon 
deuteration of the ethylene groups of the organic
molecule (BEDT-TTF). The origin of this effect is not understood, but 
differences in Fermi surface warping \cite{alessandro} or internal, uniaxial 
lattice pressure effects \cite{schlueter} have been suggested. Isotope substitution 
on any of the other atoms in the organic molecule results in a small positive 
or no isotope effect \cite{kini}. The pressure dependence of T$_{\rm c}$ in
$\kappa$-(BEDT-TTF)$_{2}$Cu(SCN)$_{2}$, however, is not affected by the 
isotope composition \cite{klehe1}, as was determined in ac-susceptibility
measurements where helium was used as a pressure medium for all samples. 
The latter result highlights that the
difference in pressure dependence upon deuteration seen in  \cite{tim}
does not originate in the isotope composition, but that
other experimental factors, such as the choice of pressure medium, might be relevant.

The main aim of this publication is to compare magnetoresistive measurements under
pressure on $\kappa$-(BEDT-TTF)$_{2}$Cu(SCN)$_{2}$ using various pressure transmitting
media, namely  
helium, Fluorinert \cite{tim}
or petroleum spirit \cite{caulfield1,caulfield2}.
Results obtained with Fluorinert and petroleum spirit differ from those
with helium and with each other. This comparison highlights that 
the experimentally determined pressure is not a transferable parameter. 
When analyzing the relationship between parameters determined 
simultaneously in each pressure experiment, we find that T$_{\rm c}$
exhibits a unique dependence on the 
Q2D carrier density in the material, 
independent of the pressure medium used. 
The increase of the Q2D carrier density can be
understood as a charge carrier transfer from the Q1D sections 
of the Fermi surface to the Q2D hole pockets.
Assuming that it is only the Q2D carriers that are 
superconducting \cite{Louati}, pressure applied to
$\kappa$-(BEDT-TTF)$_2$Cu(SCN)$_{2}$ can be understood to increase the carrier
concentration of the superconducting carriers, similar to the pressure-induced
carrier increase observed in cuprate superconductors.

\section{Experiment}

Single crystals of $\kappa $-(d$_{8}-$BEDT-TTF)$_{2}$Cu(SCN)$_{2}$ 
were grown by standard electro-crystallization
techniques \cite{kini}. The sample of $\kappa$-(d$_8$-BEDT-TTF)$_{2}$Cu(SCN)$_{2}$ 
investigated is from the same batch as that investigated in \ \cite{klehe1}.
Magnetotransport measurements under pressure on the organic superconductor
$\kappa$-(d$_{8}$-BEDT-TTF)$_{2}$Cu(SCN)$_{2}$ were executed in a 15T
superconducting magnet from Oxford Instruments. The pressure cell used was a 
gas-pressure cell from Unipress, attached to a Harwood Engineering compressor 
system with helium as the pressure medium. Above the melting curve of helium, the pressure in 
the cell is 
measured with a calibrated, temperature-compensated Manganin gauge mounted in 
the pressure circuit at room temperature. The pressure in solid helium is calculated
using the gas pressure measured and isochoric data \cite{spain}. 
During operation the pressure cell 
is permanently connected to a pressurized gas reservoir at room temperature via a 
CuBe-capillary, thus reducing the pressure decrease in the cell during the cool 
down of the cell from room temperature to 1.5 K, the base temperature of our 
experiment.

 The pressure cell is placed 
 in the centre of the magnet bore inside a variable temperature insert. The 
 temperature of the cell is regulated by varying the temperature of 
 the variable temperature insert and by adjusting the power 
 on a 2-part heater placed directly on the outer cell-body. 
 Only the top part of the cell
 heater was heated during the transition of the pressure cell 
 through the melting curve of helium, to artificially increase the temperature
 gradient over the pressure cell and thus to reduce
 strain in the sample \cite{schirber1}. 
 
 Comparison of ambient pressure Shubnikov-de Haas frequency \cite{john1} to
 that measured on the crystal mounted inside the pressure cell revealed that the
 latter was aligned with its out-of-plane axis $\sim$ 21 $^{\small{\circ}}$ off 
 the magnetic field direction and corrections to the measured Shubnikov-de Haas
 frequencies were made accordingly \cite{john1}. The resistance of the crystal 
 was measured in the inter-plane direction with a
 current of $I=9$ $\mu$A at a frequency of 33 Hz. The minimum temperature 
 accessible in these measurements was 1.5 K and the maximum magnetic field was 15 T.
 The temperature of the sample was determined with a RuO
 thermometer thermally anchored to the top of the pressure cell. 
 
 Details of the measurements on $\kappa$-(d$_8$-BEDT-TTF)$_{2}$Cu(SCN)$_{2}$ with
 Fluorinert as a pressure medium can be found in  \cite{tim}. The 
 corresponding 
 measurements on $\kappa$-(h$_8$-BEDT-TTF)$_{2}$Cu(SCN)$_{2}$ with petroleum spirit
 are described in
 detail in  \cite{caulfield1,caulfield2}.

\section{Experimental Results}
\begin{figure*}[bth]
\centering \includegraphics[height=9.1cm]{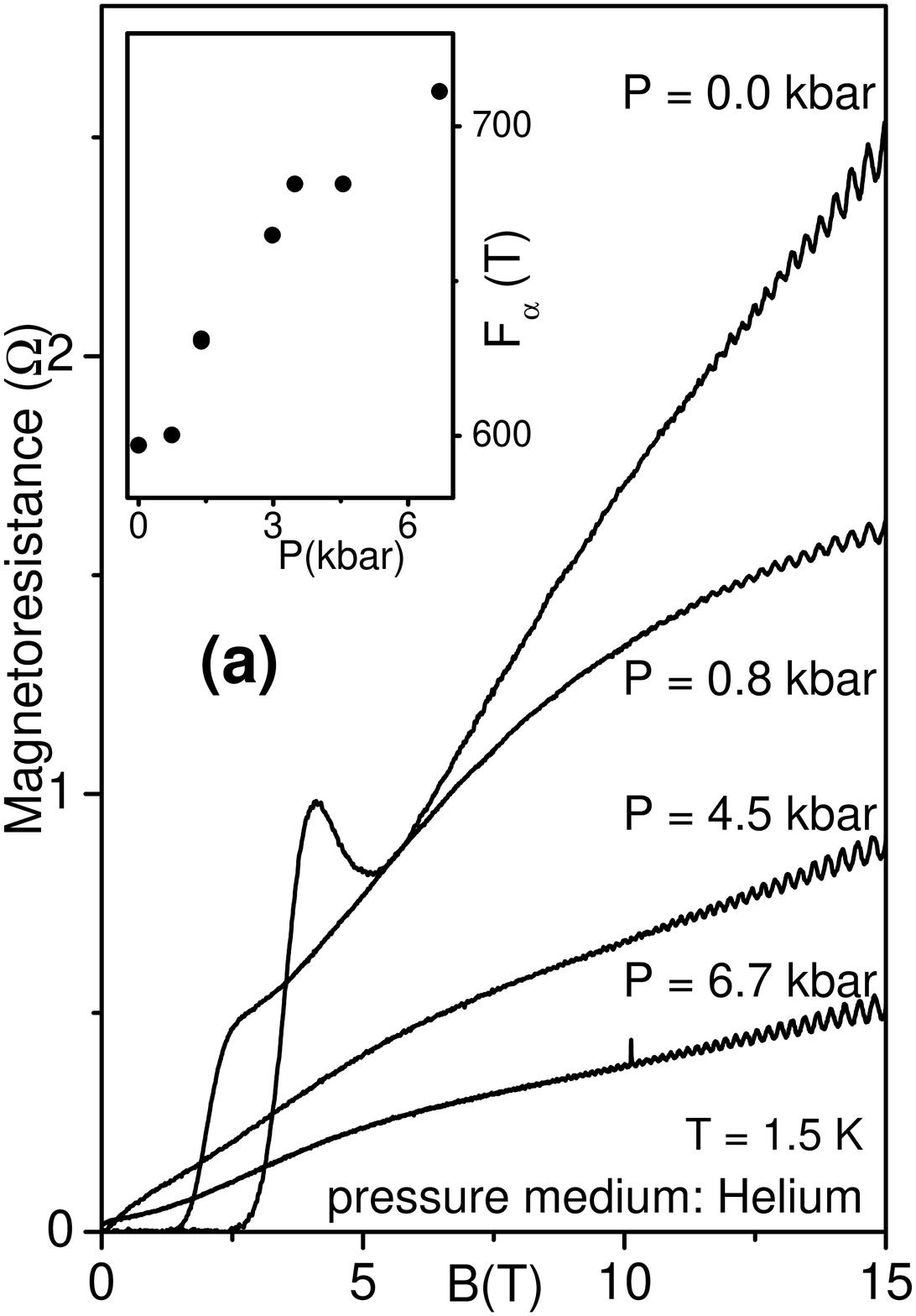}
\centering \includegraphics[height=9.1cm]{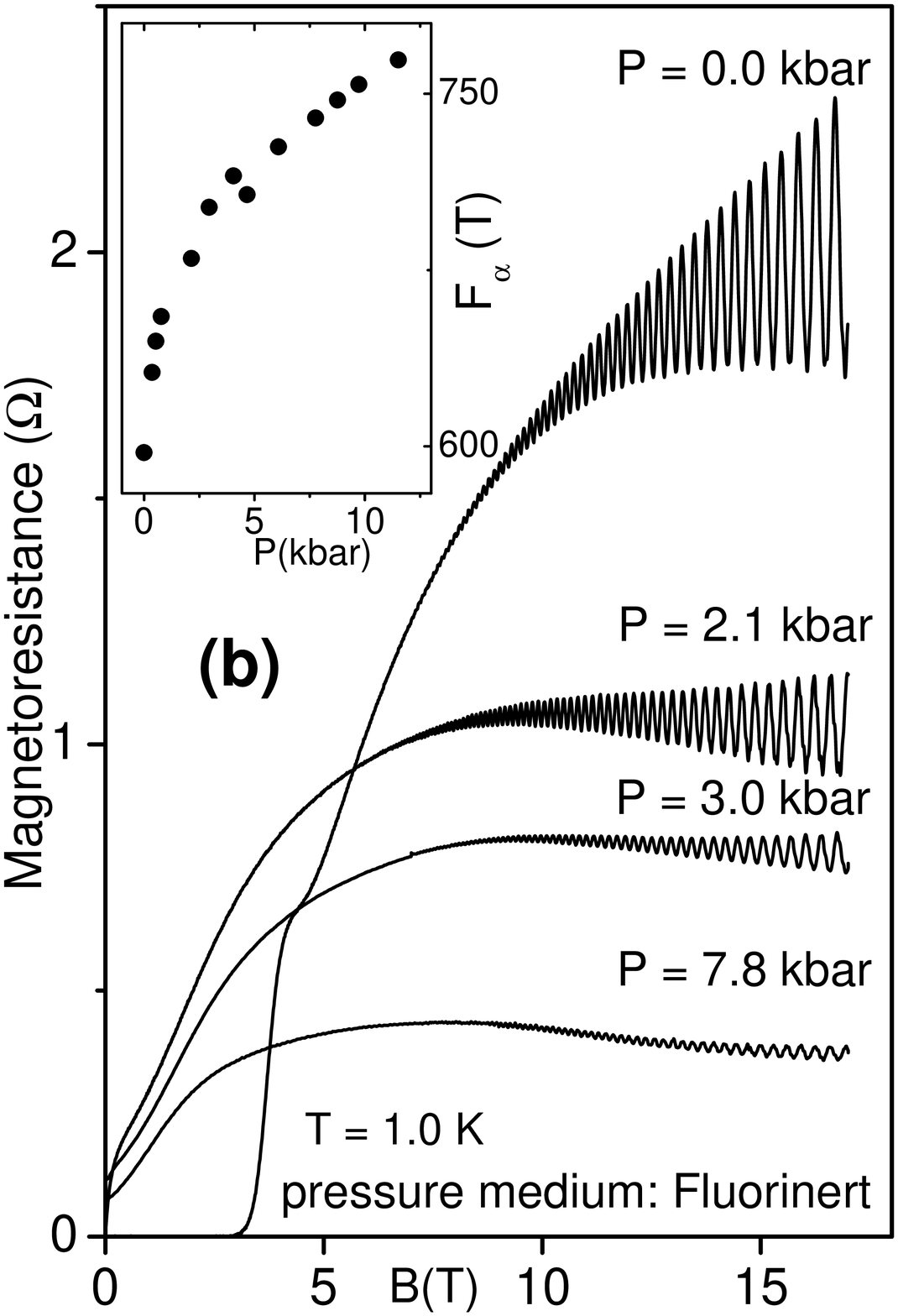}
\caption{(a) The magnetoresistance, exhibiting quantum oscillations, 
of $\kappa$-(d$_{8}-$BEDT-TTF)$_{2}$Cu(SCN)$_{2}$
at T$=1.5$ $\mathrm{K}$ at selected pressures with helium as a pressure medium. 
At T$=1.5$ $\mathrm{K}$, superconductivity at zero field is not suppressed until 
the pressure exceeds $4.5$ kbar. The insert shows the increase of the $\alpha$-
frequency, F${_\alpha}(T)$, with pressure. 
(b) The magnetoresistance of the same material, but using Fluorinert as a
pressure medium \cite{tim}. In this pressure medium, superconductivity is 
suppressed  below $T=1.0$ $\mathrm{K}$ at pressures exceeding $2.1$ kbar. 
The insert shows again the increase of F${_\alpha}(T)$ with pressure.
}
\label{magres1}
\end{figure*}

Figure\ \ref{magres1}(a) shows the magnetoresistance of 
$\kappa$-(d$_{8}-$BEDT-TTF)$_{2}$Cu(SCN)$_{2}$ at selected pressures using helium
as a pressure medium. Clearly visible are quantum oscillations associated with
the $\alpha$-frequency. 
The pressure dependence of these oscillations is illustrated in the
insert of figure \ref{magres1}(a). As is known from several previous measurements 
\cite{tim,caulfield2} the size of the Q2D Fermi surface
pocket increases with increasing pressure. The relatively high temperatures 
(T $\geq 1.5$ K) and the relatively small magnetic fields (B $\leq 15$ T) used 
in these measurements prevented us from observing the $\beta$-frequency 
(see section \ref{sec:level1}). At low pressures
one can also see the well-known peak in the magnetoresistance, thought to originate from
the motion of flux-lines in the magnetic field \cite{john1}. Superconductivity
in this measurement is seen to prevail above $1.5$ K at pressures up to $4.5$ kbar.

In comparison, in figure \ref{magres1}(b) the magnetoresistance under pressure of
$\kappa$-(d$_{8}-$BEDT-TTF)$_{2}$Cu(SCN)$_{2}$ with Fluorinert
as a pressure medium is shown. The quantum oscillations visible in these measurements are
a combination of those originating from the Q2D $\alpha$-pocket
and the magnetic breakdown $\beta$-orbit. The insert in 
figure \ref{magres1}(b) shows the development of the $\alpha$-frequency with pressure.
Details of those measurements can be found in \cite{tim};
they show that with Fluorinert as the 
pressure medium T$_{\rm c}$ is suppressed below $T=1$ ${\rm K}$ at pressures 
exceeding $2.1$ kbar. This value is much smaller than that obtained from our 
measurements with
helium as a pressure medium. 

\section{Discussion}

\subsection{The effect of the pressure medium}
\begin{figure}[thb]
\centering \includegraphics[height=6cm]{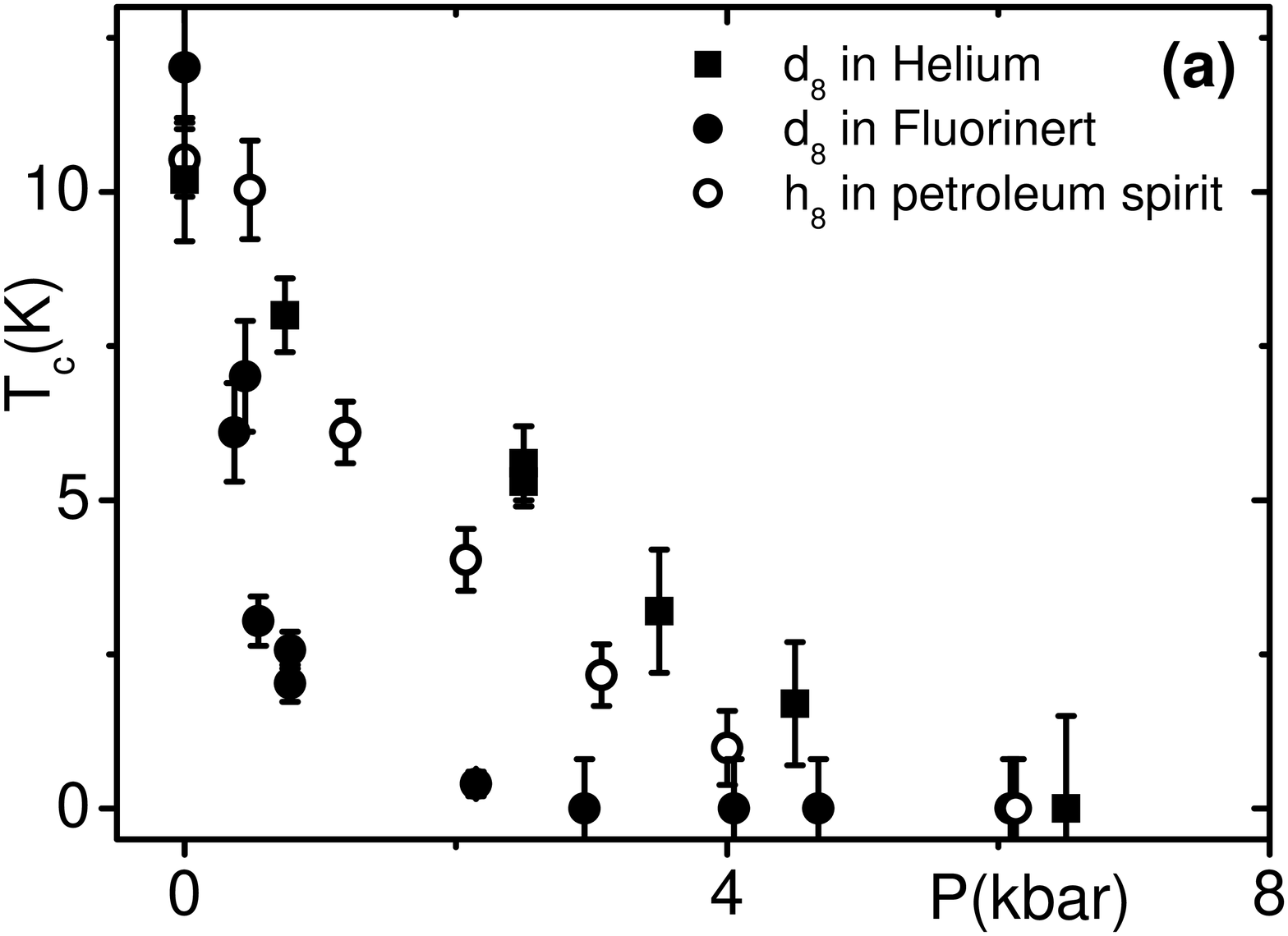}
\centering \includegraphics[height=6cm]{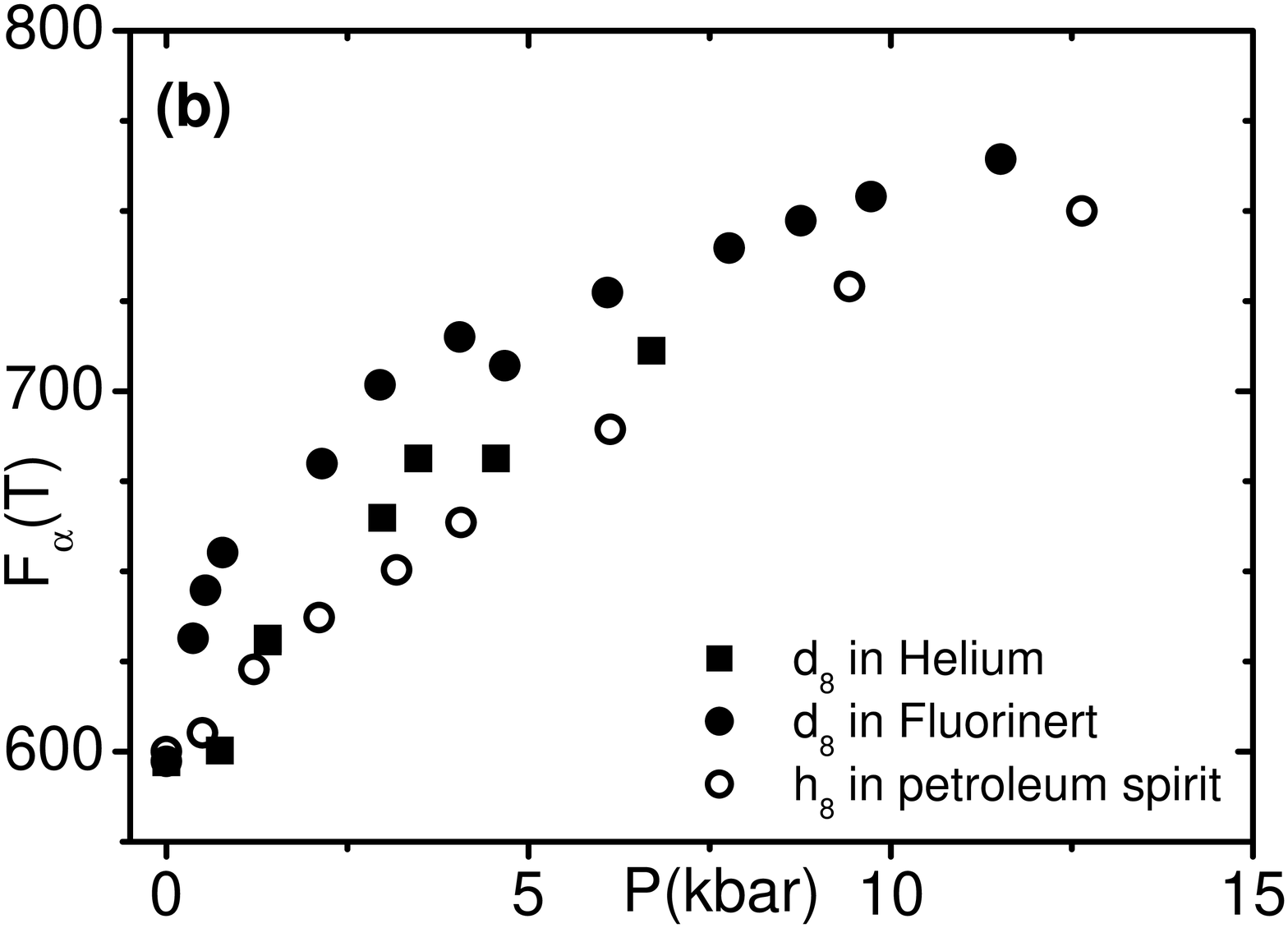}
\caption{(a) T$_{\rm c}$ and (b) $\alpha$-frequency, F$_{\alpha}$(T), as a function of pressure 
for 
$\kappa$-(d$_{8}-$BEDT-TTF)$_{2}$Cu(SCN)$_{2}$ using helium ({\tiny{$\blacksquare$}}),
using Fluorinert ({\large{\textbullet}}) \cite{tim} and for 
$\kappa$-(h$_{8}-$BEDT-TTF)$_{2}$Cu(SCN)$_{2}$ using petroleum spirit 
({\large{$\circ$}}) \cite{caulfield2} as a pressure
medium .
}
\label{tc-p}
\end{figure}

Isotopic substitution does not affect
the pressure dependence of T$_{\rm c}$
\cite{klehe1}. Thus, the magnetoresitance data under pressure on 
$\kappa$-(h$_8$-BEDT-TTF)$_{2}$Cu(SCN)$_{2}$ \cite{caulfield1,caulfield2} using 
petroleum spirit as a pressure medium are included in the comparison seen in figure
\ref{tc-p}.
Figure \ref{tc-p} compares measurements with three 
different pressure media: helium, 
Fluorinert and petroleum spirit. The three measurements indicate 
three different pressure 
dependences and only
agree in their general trends: i) T$_{\rm c}$ is decreasing 
and ii) the $\alpha$-frequency, F$_{\alpha}$(T), is 
increasing with increasing pressure.  
Having ruled out that the isotope composition \cite{klehe1} or the
source of the sample \cite{tim} could affect the pressure dependence, 
one has to conclude that the pressure medium has to be responsible for the
difference in pressure dependence observed in
$\kappa$-(BEDT-TTF)$_{2}$Cu(SCN)$_{2}$. 
No evidence for shear stresses on 
the sample could be observed in any of those measurements,
{\it i.e.} the superconducting
transitions did not broaden under pressure nor could a reduction of the 
amplitude of quantum oscillations as a function of pressure history be observed.
There was also no evidence in any of those measurements of any sample 
deterioration
after the pressure measurements, which could have indicated a chemical 
reaction between sample
and pressure medium.

The pressure media were selected because 
of their quasi-hydrostatic and hydrostatic properties, but all 
will have solidified at the temperatures at which most T$_{\rm c}$ and 
quantum oscillation measurements were executed. Thus, in most measurements, the 
sample investigated is submerged in a pressurized solid during the measurement. 
The pressure felt by the sample depends on the difference
in the thermal contractions of the pressure cell, the pressure medium and the 
sample itself \cite{schirber1}. $\kappa$-(BEDT-TTF)$_{2}$Cu(SCN)$_{2}$ is a 
very soft \cite{rahal} material with 
a highly anisotropic thermal contraction 
\cite{schultz1}. The difference in its anisotropic thermal contraction from that 
of the uniformly
contracting pressure medium results in non-hydrostatic stresses on the sample
upon cooling. Evidence for this effect of non-hydrostatic, 
pressure-medium-induced stresses on a soft sample
with an anisotropic compressibility \cite{steinle-neumann} are well documented for 
zinc \cite{schirber1,takemura}. Thus, similar effects of a frozen pressure-medium on
the soft and  highly anisotropic \cite{rahal} 
$\kappa$-(BEDT-TTF)$_{2}$Cu(SCN)$_{2}$ 
are not surprising. 

Independent measurements in different pressure cells 
by different groups \cite{tim,agosta} with 
Fluorinert as a pressure medium give the same pressure dependence of 
T$_{\rm c}$. This indicates that 
the non-hydrostatic pressure conditions caused by the 
difference in thermal contraction between a very soft, anisotropic sample 
and the pressure medium can be reproducible.
Also measurements of dT$_{\rm c}$/dP on 
$\kappa$-(BEDT-TTF)$_{2}$Cu(SCN)$_{2}$ using helium as a pressure medium give,
within the error bars, identical results, as can be
seen by comparing the results obtained in  \cite{klehe1} and those 
of the present measurement, but they are different
from those obtained with the other pressure media 
(see figure \ref{tc-p}).

Helium is generally considered to be the most hydrostatic and the softest 
pressure medium, even when it is solidified under pressure 
\cite{spain,schirber1}. At $P=1$ kbar, solid helium has a bulk modulus, 
B$_{\rm solid \hspace{.2em} helium}
\sim 6$ kbar \cite{spain}, compared to an estimated 
B$_{\rm sample} \sim 123$ kbar \cite{rahal} for 
$\kappa$-(BEDT-TTF)$_{2}$Cu(SCN)$_{2}$.
One 
could thus argue that those measurements with helium as a pressure medium reflect
the most accurate hydrostatic pressure dependence of 
$\kappa$-(BEDT-TTF)$_{2}$Cu(SCN)$_{2}$. 
Helium, on the other hand, has been known to penetrate the structure of 
some materials upon use as a pressure medium \cite{C60}. Recent experiments 
\cite{klehe1}, however,  
indicated that helium penetration into the structure of 
$\kappa$-(BEDT-TTF)$_{2}$Cu(SCN)$_{2}$ under pressure is unlikely, due to the 
insensitivity of T$_{\rm c}$ to the temperature of the pressure change, even though 
the possibility of intercalation cannot be excluded.

In summary, it appears from figure \ref{tc-p} and the above discussion
that the experimentally determined pressure cannot be taken as a reliable 
parameter to describe
the physical properties of $\kappa $-(BEDT-TTF)$_{2}$Cu(SCN)$_{2}$. This is 
most likely due to the fact that fully hydrostatic conditions of the 
pressure medium at low temperatures 
cannot be guaranteed.

\subsection{The pressure medium independent correlation between T$_{\rm c}$ and 
the quasi two-dimensional carrier density, n$_{\rm Q2D}$}

\begin{figure}[tb]
\centering \includegraphics[height=8cm]{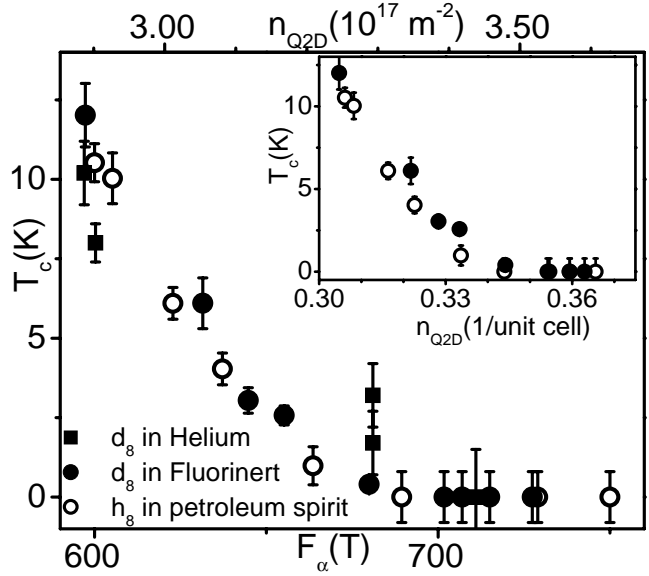}
\caption{T$_{\rm c}$ and F$_{\alpha}$ from figure \ref{tc-p}, 
indicating their correlation. The correlation is independent of 
the individual pressure dependence observed, and thus of the individual 
pressure scale. The insert shows T$_{\rm c}$ against the quasi 
two-dimensional carrier density per unit cell, 
n$_{Q2D}(\frac{1}{{\rm {unit\hspace{.07cm} cell}}})$.
}\label{tc-alpha}
\end{figure}

Common to all measurements in figure \ref{tc-p} is the simultaneous
determination of T$_{\rm c}$ and F$_{\alpha}$ under pressure. 
Figure \ref{tc-alpha} shows the correlation between these two parameters. Even 
though each of the measurements in figure \ref{tc-p} exhibits its 
own, individual pressure dependence of T$_{\rm c}$ and F$_{\alpha}$, these 
parameters demonstrate a strong correlation to each other, seemingly independent
of the pressure medium used and thus independent of the possible degree of 
shear on the sample.
Thus, the strong correlation between the bulk property of 
superconductivity, as indicated by T$_{\rm c}$, and the two-dimensional 
Fermi surface parameter, F$_{\alpha}$, suggests that superconductivity in this
organic superconductor is fully determined by processes in the quasi 
2-dimensional conducting planes. Hence, warping effects of the Fermi surface in
the interplane direction \cite{tim,alessandro}
cannot be of major significance for the superconducting properties in 
$\kappa$-(BEDT-TTF)$_{2}$Cu(SCN)$_{2}$. A similar independence of the 
superconducting properties on the interlayer parameters were indicated in
 \cite{klehe1}.

Figure \ref{tc-alpha} might partially
explain the inverse isotope effect observed in 
$\kappa$-(d$_{8}-$BEDT-TTF)$_{2}$Cu(SCN)$_{2}$; the quantum oscillation frequency 
for the Q2D Fermi surface pocket is $F_{\alpha}=600 \pm 1 {\rm T}$
for h$_{8}$ and $F_{\alpha}=597 \pm 1 {\rm T}$ for d$_{8}$ samples, which even 
though very close were consistently smaller for the d$_{8}$-sample 
\cite{tim,john-priv}.
According to figure \ref{tc-alpha}, a smaller $F_{\alpha}$ leads to a higher 
T$_{\rm c}$, as is observed upon deuteration of the organic BEDT-TTF molecule. 
The origin of the decrease in F$_{\alpha}$ upon deuteration, however, is unknown.
In addition, $\kappa$-(BEDT-TTF)$_{2}$Cu[N(CN)$_{2}$]Cl with 
T${\rm c}\approx 12.7 {\rm K}$ has a Q2D Fermi surface pocket with 
$F_{\alpha}\approx 577 {\rm T}$ \cite{kartsovnik}. This is again in agreement 
with the tendency that smaller quasi
2-dimensional Fermi surface areas result in higher T${\rm c}$.  

Considering that the degree of coherent transport in the interplane direction
in $\kappa$-(BEDT-TTF)$_{2}$Cu(SCN)$_{2}$ is very small \cite{john3}, we shall in
the following treat it as a 2-dimensional, 2D, electronic system:
in a 2D electronic system, simple Fermi statistics correlates
the frequency of the 2D Fermi surface pocket, F$_{\alpha}$, to the
carrier density, n$_{\rm Q2D}$, of that pocket:

\begin{equation} \label{eq1}
n_{\rm Q2D}({\rm m^{-2}}) = \frac{A_{\rm FS}}{2 \pi^{2}}
\end{equation}
with
\[
A_{\rm FS} = \frac{2 \pi {\rm e}}{\hbar}F_{\alpha}.
\]

The observed correlation between T$_{\rm c}$ and F$_{\alpha}$ is thus synonymous
with a correlation between  T$_{\rm c}$ and the Q2D carrier 
density, n$_{\rm Q2D}$. 
n$_{\rm Q2D}$ which has been calculated according to equation
(\ref{eq1}) is plotted as the top x-axis in Fig. \ref{tc-alpha}: 
the effect of pressure is to increase the number of carriers 
in the Q2D hole pockets in 
$\kappa$-(BEDT-TTF)$_{2}$Cu(SCN)$_{2}$. 
For the measurements in Fluorinert \cite{tim} and petroleum spirit \cite{caulfield1}
the in-plane compression of the unit-cell at low temperatures has been determined
from the pressure dependence of the $\beta$-frequency to be $\sim$ 4\%/GPa.
This compares to an overall increase in F$_\alpha$ or n$_{Q2D}$(m$^{-2}$)
of $\sim$ 30\% in the same pressure region, 
indicating that n$_{Q2D}$(m$^{-2}$) increases beyond what would be 
expected from the compression of the Brillouin zone alone. 
Given the in-plane compressibility information \cite{tim,caulfield1}  
n$_{Q2D}$(m$^{-2}$) for the Fluorinert and petroleum spirit measurement
is converted to the carrier density-per-unit-cell,
n$_{Q2D}$(1/unit\hspace{.2em}cell).
It appears from the insert of Fig. \ref{tc-alpha} that 
n$_{Q2D}$(1/unit\hspace{.2em}cell) has 
increased from $\sim$ 0.30 (holes/unit\hspace{.2em}cell) to 
$\sim$ 0.34 (holes/unit\hspace{.2em}cell) when T$_{\rm c}$ is suppressed.
Thus, in the temperature-pressure phase 
diagram of $\kappa$-(BEDT-TTF)$_{2}$Cu(SCN)$_{2}$ one can think of  
pressure as a driving force for increasing the Q2D carrier 
density, n$_{Q2D}(m^{-2})$.
However, whereas experimentally-determined pressure is in general not a 
suitable parameter to describe
the superconducting state of the system (see figure \ref{tc-p}), the carrier 
density, n$_{Q2D}$, obtained from F$_{\alpha}$ and the in-plane compressibility,
predicts 
T$_{\rm c}$ extremely well.

$\kappa$-(BEDT-TTF)$_{2}$Cu(SCN)$_{2}$ has two holes per unit cell
\cite{yamaji} due to the transfer of two electrons from the 4 (BEDT-TTF)-molecules
per unit cell to the polymorphic Cu(SCN)$_{2}$-layer. 
At the small pressures used in our experiments, this overall
carrier density per unit cell has to be considered independent of pressure. 
In the conducting organic layer in $\kappa$-(BEDT-TTF)$_{2}$Cu(SCN)$_{2}$, 
these holes are distributed between 
the Q2D Fermi surface pockets and the Q1D 
Fermi surface sheets. 

\begin{centering}
\begin{eqnarray}\label{ntotal}
n_{total}(1/{\rm unit}\hspace{.2em}{\rm cell}) & \equiv & 
				 2 \hspace{.2em}{\rm holes}/{\rm unit}\hspace{.2em}{\rm cell} \\
 				 & = & n_{Q2D}(1/{\rm unit}\hspace{.2em}{\rm cell}) \nonumber \\
				 & & + n_{Q1D}(1/{\rm unit}\hspace{.2em}{\rm cell}) \nonumber 
\end{eqnarray}
\end{centering}

According to equation (\ref{ntotal}) the observed increase in n$_{Q2D}$
is equivalent to an identical decrease in n$_{Q1D}$. Thus, the increase in  
n$_{Q2D}$ can be understood as a simple, pressure-induced charge transfer of 
holes from the
quasi one-dimensional Fermi surface planes to the quasi two-dimensional 
Fermi surface pockets in $\kappa$-(BEDT-TTF)$_{2}$Cu(SCN)$_{2}$.
Based on our experiments, it is not possible to decide whether 
it is the increase 
in $n_{Q2D}$ or the decrease in $n_{Q1D}$ that is the relevant parameter for 
the suppression of T$_{\rm c}$. 

\begin{figure}[tb]
\centering \includegraphics[height=6cm]{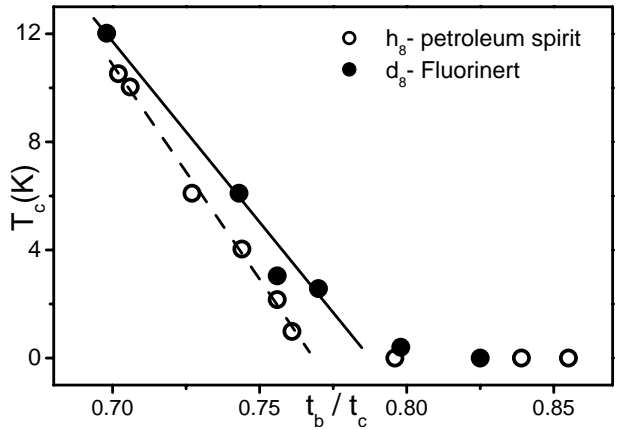}
\caption{T$_{\rm c}$ as a function of the ratio of the in-plane transfer 
integrals, t$_{b}$/t$_{c}$ \cite{tim}. A similar trend
can be observed in both measurements but there is no strong correlation between 
T$_{\rm c}$ and 
t$_{b}$/t$_{c}$.
}\label{Tc-tbtc}
\end{figure}

This mechanism bears a resemblance to the mechanism observed in 
cuprate superconductors where the effect
of pressure is to transfer holes from the insulating charge reservoir layers 
into the Q2D, conducting CuO$_{2}$ layers \cite{neumeier}. 
Thus, in organic 
superconductors,
similar to cuprate superconductors, pressure has the effect of increasing the 
Q2D carrier density by transfering holes from other parts of the
Fermi surface to those bands that support superconductivity \cite{Louati}.

Fig. \ref{tc-alpha} demonstrates very clearly that there is a strong dependence
between T$_{\rm c}$ and n$_{Q2D}$, which in turn is intimately related to n$_{Q1D}$.
The sum of those two parameters is always equal to two holes per in-plane unit 
cell area. Thus, it is the in-plane pressure that seems to be relevant for the 
superconducting properties in $\kappa$-(BEDT-TTF)$_{2}$Cu(SCN)$_{2}$ and not the
bulk pressure, as would be indicated by the pressure gauge. Such a dependence
of the superconducting properties on the compression of the in-plane unit cell 
size has been suggested 
by recent a.c.-susceptibility measurements under pressure \cite{klehe1}.

Louati {\it et al.} \cite{Louati} considered the ratio of the transfer integrals
t$_{b}$/t$_{c}$ to be the most critical parameter for $\kappa$-(BEDT-TTF)$_{2}$X
salts. This ratio of transfer integrals has been calculated for the Fluorinert 
and petroleum spirit measurements \cite{tim}, and is plotted against T$_{\rm c}$
in figure \ref{Tc-tbtc}. Both measurements exhibit a similar trend in the dependence
of T$_{\rm c}$ on this ratio, but the exact dependence is seen to be affected
by the pressure medium (see figure \ref{Tc-tbtc}). Also the correlation 
between T$_{\rm c}$ and the effective mass
is known to exhibit a dependence on the pressure medium \cite{tim}.
In contrast, no dependence on the pressure medium could be seen
in the correlation between T$_{\rm c}$ and n$_{Q2D}$ 
(see figure \ref{tc-alpha}), indicating that the superconducting carrier density 
could be a major determining factor for the superconducting properties of the organic 
superconductor $\kappa$-(BEDT-TTF)$_{2}$Cu(SCN)$_{2}$.

\section{Conclusion}
We compared different magnetotransport measurements under pressure on the 
organic superconductor $\kappa$-(BEDT-TTF)$_{2}$Cu(SCN)$_{2}$ and found that 
pressure in general is not a reliable parameter for this system. This is 
thought to be due to the strong anisotropy of all physical properties in 
$\kappa$-(BEDT-TTF)$_{2}$Cu(SCN)$_{2}$ and possible departure from fully hydrostatic
conditions when the sample is cooled in the frozen pressure medium.
However, a strong correlation between
the superconducting transition temperature, T$_{\rm c}$, and the carrier density
of the Q2D Fermi surface, n$_{Q2D}$, can been observed (see figure (\ref{tc-alpha})). This 
correlation is independent of the pressure medium used. 

Pressure can be understood to increase n$_{Q2D}$ by transferring carriers from 
the Q1D sections of the Fermi surface to the Q2D
sections. Thus the effect of pressure on an organic superconductor resembles that
observed in cuprate superconductors: in both materials pressure causes the transfer 
of holes from non-superconducting sections of the Fermi surface to the superconducting sections.

\ack{
Research at the Clarendon Laboratory, Oxford University, is supported by the EPSRC, 
Grant No. GR/R16075/01. Research at Argonne National Laboratory is sponsored by 
the US Department of
Energy, Office of Basic Energy Sciences, Division of Materials Sciences
under contract W-31-109-ENG-38. C.\ A.\ K.\ thanks the Deutsche Forschungsgemeinschaft
for financial support.
Special thanks are due to J.\ Annett for his stimulating discussion. The authors
thank P.\ A.\ Goddard, S.\ Blundell and W.\ Hayes for their assistance in 
the preparation of
this manuscript.}

\end{document}